\documentclass[12pt]{iopart}
\usepackage{iopams}

\expandafter\let\csname equation*\endcsname\relax

\expandafter\let\csname endequation*\endcsname\relax

\usepackage{graphicx,color} 

\usepackage{amsmath,amsfonts,amssymb,theorem}
\usepackage{url,subfig,anysize,wrapfig}
\usepackage{multirow,tabularx,afterpage,fancyhdr}
\usepackage{cases}
\usepackage{epstopdf}
\usepackage{bm}

\usepackage{lipsum}

\usepackage{lineno}
\usepackage{placeins}

\usepackage{tikz}

\begin{document}

\title[]{Transition criteria and phase space structures in a three degree of freedom  system with dissipation}

\author{Jun Zhong \textsuperscript{1} \& Shane D. Ross \textsuperscript{2}}

\address{$^1$School of Engineering, Brown University, Providence, RI 02912, USA\\
$^2$Aerospace and Ocean Engineering, Virginia Tech, Blacksburg, VA 24061, USA}
\ead{jun\_zhong@brown.edu (J.Zhong), sdross@vt.edu (S.D. Ross)}

\vspace{10pt}
\begin{indented}
\item[] 1 June 2021
\end{indented}

\begin{abstract}
Escape from a potential well through an index-1 saddle can be widely found in some important physical systems. Knowing the criteria and phase space geometry that govern  escape events plays an important role in making use of such phenomenon, particularly when realistic frictional or dissipative forces are present.
We aim to extend the study the escape dynamics around the saddle from two degrees of freedom to  three degrees of freedom, presenting both a methodology and phase space structures. 
Both the ideal conservative system and a perturbed, dissipative system are considered. 
We define the five-dimensional transition region, $\mathcal{T}_h$, as the set of initial conditions of a given initial energy $h$ for which the trajectories  will escape from one side of the saddle to another. Invariant manifold arguments demonstrate that in the six-dimensional phase space, the boundary of the transition region, $\partial \mathcal{T}_h$, is topologically  a four-dimensional hyper-cylinder in the conservative system, and  a four-dimensional hyper-sphere in the dissipative system.  
The transition region 
$\mathcal{T}_h$ can be constructed by a solid three-dimensional ellipsoid (solid three-dimensional cylinder) in the three-dimensional configuration space, where at each point, there is a cone of velocity---the  velocity directions leading to transition are given by cones, with velocity magnitude given by the initial energy and the direction by two spherical angles with given limits. 
To illustrate our analysis, we consider an example system which has two potential minima connected by an index 1 saddle.
\end{abstract}

%
\vspace{2pc}

\section{Introduction}
From the perspective of multi-degree of freedom mechanical systems, transition events can be interpreted as the escape from potential wells. They are widely found in a number of important physical systems, such as snap-through buckling of curved structures \cite{zhong2021differential,virgin2017geometric,napoli2015snap}, ship motion and capsize \cite{NaRo2017,nayfeh1973nonlinear}, chemical reactions \cite{uzer2002geometry,feldmaier2019invariant}, and celestial mechanics \cite{jaffe2002statistical,KoLoMaRo2000}. Good understanding of the dynamical structures and transition criteria can promote engineering design. Due to the importance of accurately predicting the transition events, a plenty of study has been conducted, presenting both computational algorithms and phase space structures.

For conservative systems with higher degrees of freedom, the transition boundary for all possible escape trajectories is known to be the stable and unstable invariant manifolds of a normally hyperbolic invariant manifold (NHIM) \cite{Wiggins1994} of a given energy. Recently, it was found that the corresponding transition boundary for dissipative systems is the stable invariant manifolds of an index-1 saddle connecting the potential wells \cite{zhong2020geometry,zhong2021global}. Specifically, for an intermediate case of two degree of freedom, the NHIM is a collection of periodic orbits around the index-1 saddle, each periodic orbit corresponding to a given energy \cite{KoLoMaRo2011,NaRo2017,lyu2020role,katsanikas2020phase}. The geometry of corresponding stable invariant manifolds of the NHIM and the index-1 saddle are pieces of tubes (or cylinders) \cite{NaRo2017,KoLoMaRo2000,naik2019finding} and ellipsoids \cite{garcia2021painting,zhong2021global,zhong2020geometry} in the conservative and dissipative systems, respectively, sometimes referred to as transition tubes and transition ellipsoids  in \textit{tube dynamics} \cite{GaKoMaRoYa2006,OnYoRo2017,zhong2018tube}. Recently, the theory of transition tubes in  two degrees of freedom  was experimentally verified \cite{Ross2018experimental}. The transition boundary or the invariant manifold of a given energy separate two types of trajectories: transit orbits that escape from one potential well to another have initial conditions inside of the boundary, while non-transit orbits that evolve within the potential well have initial conditions outside of the boundary.

After knowing the fact that the transition is governed by the invariant manifolds, the candidate idea of finding the transition boundary can be computing the invariant manifolds. For conservative systems, one needs to first compute the NHIMs and then globalize the corresponding invariant manifolds \cite{KoLoMaRo2011}. On the other hand, for dissipative systems, after determining the local invariant manifolds of the linearized systems, the global invariant manifolds can be grown from the local ones \cite{krauskopf2003computing,krauskopf2007numerical}. From the perspective of invariant manifold theory, some standard methods were proposed to compute the transition boundary in both the conservative \cite{KoLoMaRo2011} and dissipative systems \cite{zhong2021global}. Note that since the invariant manifolds for the conservative and dissipative systems are mathematically different, generally the algorithms proposed for the conservative systems can not work for the dissipative systems and vise versa.  Some methods which do not rely on any information of the invariant manifolds were also developed which work for both conservative and dissipative systems, such as Lagrangian descriptor \cite{garcia2020exploring,garcia2020tilting}, bisection method \cite{zhong2018tube}, isolating blocks \cite{anderson2017isolating,anderson2020computing}, to name but a few. Other computational algorithms can also be found in \cite{dellnitz1997subdivision,dellnitz1996computation,krauskopf2006survey,onozaki2017tube,branicki2009adaptive,castelli2015parameterization}. Compared to the well understood escape dynamics in the conservative systems with higher degrees of freedom \cite{krajvnak2021reactive,GoKoLoMaMaRo2004,naik2019finding}, we have little understanding on higher degree of freedom systems with dissipation. Here we aim to extend the study on the linearized dynamics in two degree of freedom systems \cite{zhong2020geometry} to three degree of freedom systems in the presence of dissipative forces.

In the current study, we present a construction, building off of Conley \cite{Conley1968}, to establish the criteria for initial conditions leading to transitions in a three degree of freedom dissipative system. 
We consider an exemplar canonical Hamiltonian system with three degrees of freedom which admits an index-1 saddle point will be considered, both  with and without linear dissipation. 
This system, which has two potential minima connected by an index-1 saddle, is chosen for its simplicity to illustrate our analysis. 
It resembles an isomerization system with two conformations, a three degree of freedom generalization of those  in, e.g., \cite{deleon1989order,MaDe1989,OzDeMeMa1990}.  
Application to other examples of interest is planned for future work.
By virtue of  appropriate theorems---the 
theorem of the local stable and unstable manifold \cite{meiss2007differential,wiggins2003introduction,perko2013differential} in the dissipative case,
and a theorem of Moser \cite{Moser1958,Moser1973} in the conservative case---all the qualitative results from the local linearized behavior carry over to the full nonlinear equations.

The paper is organized as follows. In Section \ref{equilibrium_equations}, we derive the equations of motion in both the Lagrangian system and Hamiltonian system. The analysis of the linearized dynamics for the conservative and dissipative systems are given in Section \ref{linearized_conservative} and Section \ref{linearized_dissipative}, respectively. The corresponding analytical solutions and topological phase space structures that govern the transition are presented. Then, in Section \ref{Discussion}, we give some discussions and remarks on the current study. Finally, we summarize the current work and discuss the possible future work in Section \ref{conclusions}.

\section{Equations of motion}
\label{equilibrium_equations}
\begin{figure}[h]
	\begin{center}
		\includegraphics[width=\textwidth]{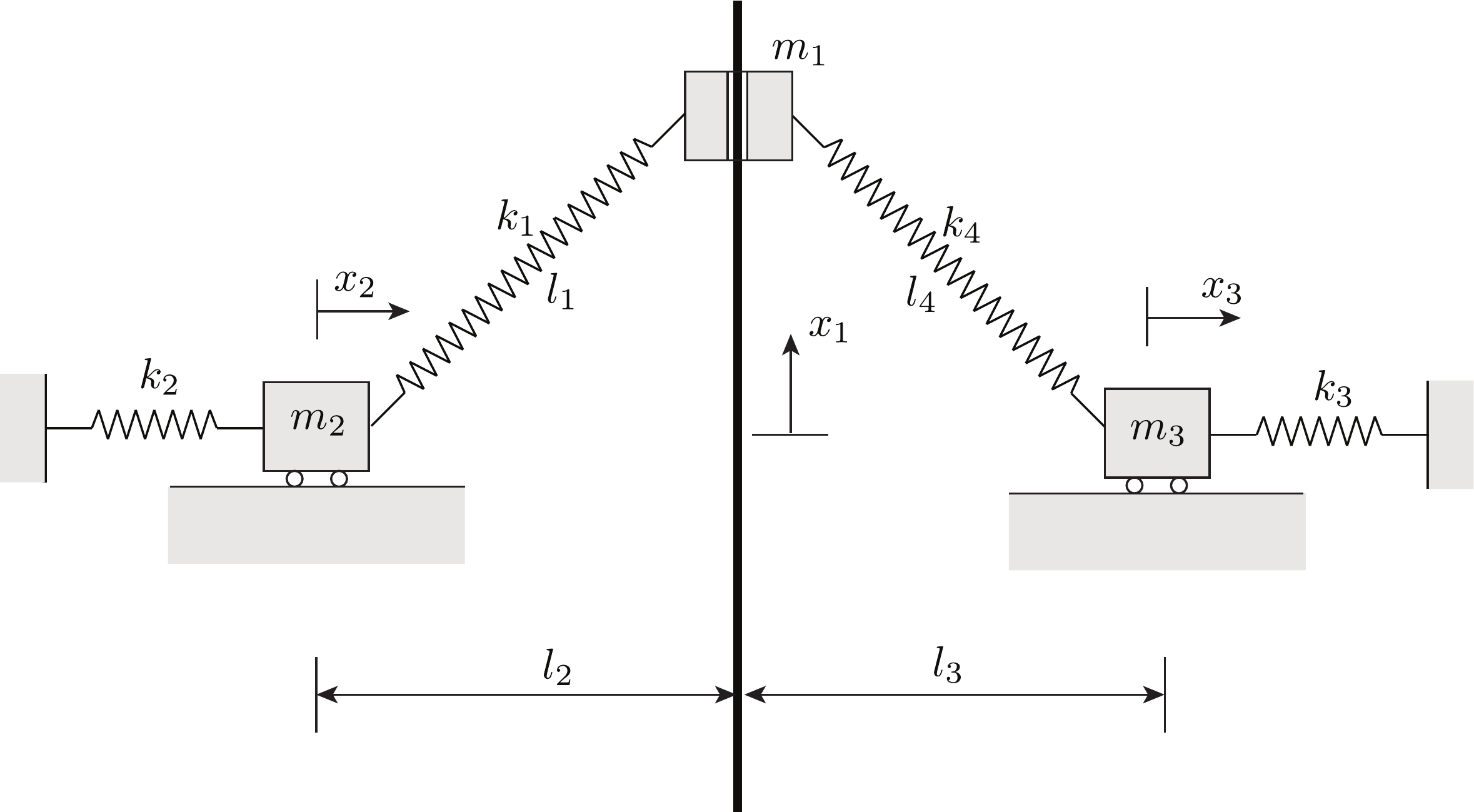}
	\end{center}
	\caption{{\footnotesize 
			Schematic of the three-degree-of-freedom spring-mass system. One of the equilibrium configurations is shown, where $x_{2e}=x_{3e}=0$, and $x_{1e}> 0$.
	}}
	\label{spring-mass schematic}
\end{figure}

Consider a planar three degree of freedom spring-mass system as shown in Figure \ref{spring-mass schematic}. 
Three masses, denoted by $m_i$ $(i=1,2,3)$, are connected by four springs with stiffnesses denoted by $k_i$ $(i=1,2,3,4)$. 
Constrained by a straight bar, $m_1$ can only slide along the vertical direction, while $m_2$ and $m_3$ can only move horizontally due to the constraint by the horizontal ground. 
Gravity is neglected.

When the system is in a stable equilibrium state, all the springs are neither compressed nor stretched. 
The equilibrium length of the spring connecting $m_1$ and $m_2$ is $l_1$ and  the equilibrium length of the spring connecting $m_1$ and $m_3$ is $l_4$. 
The horizontal distances of $m_2$ and $m_3$ to the bar at stable equilibrium are denoted by $l_2$ and $l_3$ where $l_1>l_2$ and $l_4>l_3$.
Note that the lengths $l_i$ are not independent and satisfy the geometric relation $l_1^2 - l_2^2 = l_4^2-l_3^2$. 
The generalized coordinates $\left(x_1,x_2,x_3\right)$ are established to describe the position of the masses, where the origin of the $x_1$ coordinate is at the same height of $m_2$ and $m_3$. 
Note that by symmetry, there is stable equilibrium with $x_1>0$ and a mirror image stable configuration
with $x_1<0$.
The origin of the $x_2$ and $x_3$ coordinates is at the stable configurations and both increase to the right, as shown in Figure \ref{spring-mass schematic}.
Note that there is an unstable equilibrium configuration where the masses are collinear, that is, $x_1=0$.

The equations of motion are established as follows. We initially derive the mathematical model from a Lagrangian point of view. The kinetic energy for the system is given by,
\begin{equation}
	\mathcal{T} (\dot x_1, \dot x_2, \dot x_3)=\tfrac{1}{2} \left(m_1 \dot x_1^2 + m_2 \dot x_2^2 + m_3 \dot x_3^2 \right),
\end{equation}
where the dot over a variable denotes the derivative with respect to time. The potential energy is,
\begin{equation}
	\mathcal{V} (x_1,x_2,x_3)=\tfrac{1}{2}  \left( k_1 \Delta l_1^2 + k_2 x_2^2  + k_3 x_3^2 + k_4 \Delta l_4^2 \right),
\end{equation}
where $\Delta l_1$ and $\Delta l_4$ are the changes of the length for spring 1 and spring 4, respectively, given by,
\begin{equation}
	\Delta l_1=\sqrt{\left(l_2 - x_2 \right)^2 + x_1^2 }- l_1, \hspace{0.2in} \Delta l_4=\sqrt{\left(l_3 + x_3 \right)^2 + x_1^2 }- l_4.
\end{equation}
After obtaining the kinetic energy and potential energy, the Lagrange function $\mathcal{L}$ can be defined by,
\begin{equation}
	\mathcal{L}(x_1,x_2,x_3,\dot x_1, \dot x_2, \dot x_3)= \mathcal{T}(\dot x_1, \dot x_2, \dot x_3)-\mathcal{V} (x_1,x_2,x_3),
\end{equation}
from which one can obtain the Euler-Lagrange equations,
\begin{equation}
	\frac{\mathrm{d} }{\mathrm{d} t} \frac{\partial \mathcal{L}}{\partial \dot q_i} - \frac{\partial \mathcal{L}}{\partial q_i}=Q_i, \quad i=1,2,3,
	\label{Euler_Lagrange}
\end{equation}
where $q_i$ are the generalized coordinates, and $Q_i$ are the non-conservative forces. In the current problem, we consider a small linear viscous damping, proportional to the magnitude of the inertial velocity, i.e., $Q_i = c_i \dot q_i$,  where $c_i$ $(i=1,2,3)$ are the coefficients of the viscous damping. From the Euler-Lagrange equation in \eqref{Euler_Lagrange}, the equations of motion can be finalized by,
\begin{equation}
	\begin{aligned}
		& m_1 \ddot x_1 + c_1 \dot x_1 + \frac{k_1 \Delta l_1}{\Delta l_1 + l_1} x_1 + \frac{k_4 \Delta l_4}{\Delta l_4 + l_4} x_1=0,\\
		& m_2 \ddot x_2 + c_2 \dot x_2 + k_2 x_2 + \frac{k_1 \Delta l_1}{\Delta l_1 + l_1} \left(x_2 - l_2 \right)=0,\\
		& m_3 \ddot x_3 + c_3 \dot x_3 + k_3 x_3 + \frac{k_4 \Delta l_4}{\Delta l_4 + l_4} \left(x_3 + l_3 \right)=0.
	\end{aligned}
\end{equation}
After establishing the Lagrangian system, we can transform it to a Hamiltonian system via 
the Legendre transformation,
\begin{equation}
	p_i=\frac{\partial \mathcal{L}}{\partial q_i}, \hspace{0.2in} \mathcal{H} \left(q_i,p_i \right)=\sum_{i=1}^{n} p_i \dot q_i - \mathcal{L} \left(q_i, p_i\right).
\end{equation}
Here $p_i$ is the generalized momentum conjugate to the generalized coordinate $q_i$, and $\mathcal{H}$ is the Hamiltonian function. In the spring-mass system considered here, the Legendre transformation is given by,
\begin{equation}
	\begin{aligned}
		p_1=m_1 \dot x_1, \hspace{0.2in} p_2= m_2 \dot x_2, \hspace{0.2in} p_3=m_3 \dot x_3,
	\end{aligned}
\end{equation}
and the Hamiltonian function becomes,
\begin{equation}
	\mathcal{H}= \mathcal{T}+\mathcal{V},
\end{equation}
which is the total energy, the sum of kinetic  and potential energies, where the kinetic energy 
is written in terms of the generalized momenta, 
\begin{equation}
	\mathcal{T}=\frac{1}{2} \left[\left(\frac{p_{1}}{m_1} \right)^2 + \left(\frac{p_{2}}{m_2} \right)^2 + \left(\frac{p_{3}}{m_3} \right)^2 \right].
\end{equation}
The Hamiltonian equations with damping \cite{greenwood2006advanced} are given by,
\begin{equation}
	\dot q_i=\frac{\partial \mathcal{H}}{\partial p_i}, \hspace{0.2in} \dot p_i=-\frac{\partial \mathcal{H}}{\partial q_i} + Q_i.
\end{equation}
The detailed form of the Hamiltonian equations are, 
\begin{equation}
	\begin{aligned}
		& \dot x_1= p_{1}/m_1, \hspace{0.1in}\dot x_2= p_{2}/m_2, \hspace{0.1in} \dot x_3= p_{3}/m_3,\\
		& \dot p_{1}= -\frac{k_1 \Delta l_1}{\Delta l_1 + l_1} x_1 - \frac{k_4 \Delta l_4}{\Delta l_4 + l_4} x_1 - c_1 p_{1} /m_1,\\
		& \dot p_{2}=- k_2 x_2 - \frac{k_1 \Delta l_1}{\Delta l_1 + l_1} \left(x_2 - l_2 \right) - c_2 p_{2} /m_2,\\
		& \dot p_{3}=-k_3 x_3 - \frac{k_4 \Delta l_4}{\Delta l_4 + l_4} \left(x_3 + l_3 \right) - c_3 p_{3} /m_3.
		\label{Hamilton equations}
	\end{aligned}
\end{equation}

In the current analysis, the parameters for the spring-mass system are taken as,
\begin{equation}
	\begin{aligned}
		& m_1=m_2=m_3=0.1~\mathrm{kg}, \hspace{0.1in} c_1=c_2=c_3=100~ \mathrm{N  s/m}, \hspace{0.1in} k_1=k_2=100~\mathrm{N/m},\\
		&  k_3 =k_4=150~\mathrm{N/m}, \hspace{0.1in} l_1=0.8~\mathrm{m}, \hspace{0.1in} l_2=0.5~ \mathrm{m}, \hspace{0.1in} l_3=0.6~\mathrm{m}, \hspace{0.1in} l_4=0.5\sqrt{3}~\mathrm{m}.
		\label{geometrical_parameters}
	\end{aligned}
\end{equation}
For the given parameters, the system has three physically possible equilibrium points, two stable wells and one index-1 saddle, with the following equilibrium configuration locations,
\begin{equation}
	(x_{1e},x_{2e},x_{3e})=\left\{ 
	\begin{aligned}
		&(\pm \sqrt{l_1^2-l_2^2},0,0),\hspace{0.2in} &&\text{for the stable wells},\\
		&\left(0,\frac{k_1(l_2-l_1)}{k_1+k_)},\frac{k_4(l_4-l_3)}{k_3+k_4} \right), \hspace{0.2in} && \text{for the index-1 saddle}.
	\end{aligned}
	\right.
\end{equation} 
\begin{figure}[t]
	\begin{center}
		\includegraphics[width=\textwidth]{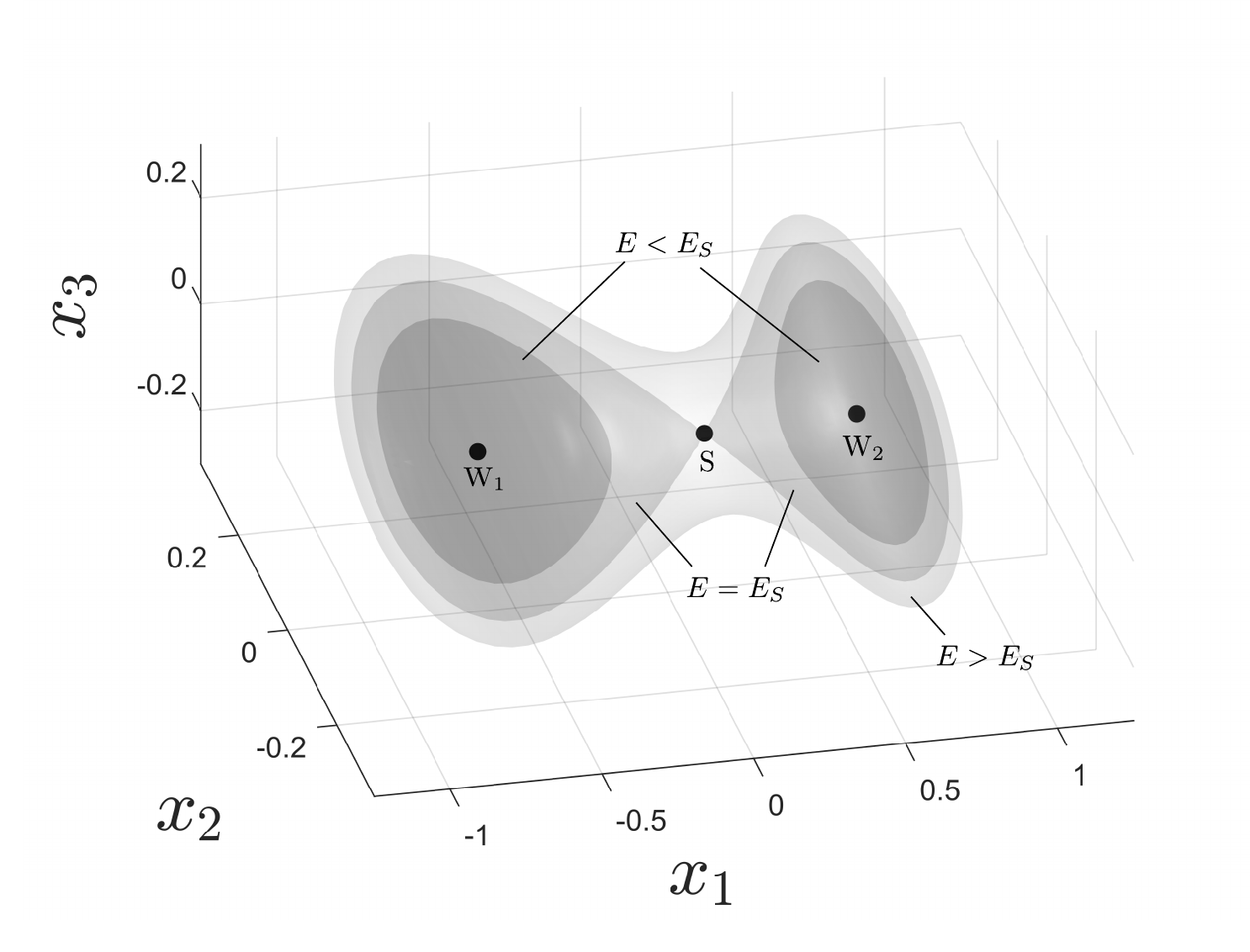}
	\end{center}
	\caption{{\footnotesize 
			Three-dimensional contour of potential energy underlying the Hamiltonian function. Here W$_1$ and W$_2$ are the stable wells, while S is the index-1 saddle. Three energy contours for $E<E_S$, $E=E_s$, and $E>E_S$ are given for shown for three different energy cases.  
	}}
	\label{energy_contour}
\end{figure}

For the parameters given in \eqref{geometrical_parameters}, three equilibria are shown in Figure \ref{energy_contour}, where W$_1$ and W$_2$ are within the two stable wells, while S denotes the index-1 saddle. Denoting the potential energy at the saddle by $E_S$, one can notice it is the \textit{critical energy} that allows the transition motion between the potential wells. The critical energy surface divides the motion into two energy cases. When the total energy is smaller than the critical energy, i.e., $E<E_s$, the motion is bounded within the potential wells, while when the total energy is bigger than the critical energy, i.e. $E>E_s$, the energy surface has a bottleneck that allows possible motion between potential wells. Three energy surface for $E<E_S$, $E=E_S$, and $E>E_S$ are also shown in Figure \ref{energy_contour}.

\paragraph{Linearization around the saddle} In the current study, we only focus on the local behavior around the index-1 saddle. Linearizing the equations of motion in \eqref{Hamilton equations} about the saddle, we use positions and momenta, 
$\left( X_1,X_2,X_3,p_{X_1},p_{X_2},p_{X_3}\right)=\left(x_1,x_2,x_3,p_{1},p_{2},p_{3} \right)-\left(x_{1e},x_{2e},x_{3e},0,0,0 \right)$,
and obtain the linearized equations as,
\begin{equation}
	\begin{aligned}
		\dot X_i= p_{X_i}/m_i, \hspace{0.1in} \dot p_{X_i}=a_i X_i - c_i p_{X_i}/m_i,  \hspace{0.2in} i=1,2,3.
	\end{aligned}
\end{equation}
By using $x_{1e}=0$, as is the case for the index-1 saddle point, it is straightforward to obtain the following coefficients,
\begin{equation}
	\begin{aligned}
		& a_1= 
		\frac{k_1 k_2 (l_1-l_2)}{k_1 l_1+k_2 l_2} + \frac{k_3 k_4 (l_4 - l_3)}{k_3 l_3 + k_4 l_4}>0,\\ 
		& a_2= -(k_1 +k_2)<0,\\ 
		& a_3=-(k_3+ k_4)<0.
	\end{aligned}
\end{equation}

For easier derivation, we introduce  scaled parameters, $P_{X_i}=p_{X_i}/m_i$, $A_i=a_i/m_i$, and $C_i=c_i/m_i$, $i=1,2,3$. 
The equations of motion can be rewritten in a simpler form,
\begin{equation}
	\begin{aligned}
		\dot X_i= P_{X_i}, \hspace{0.1in} \dot P_{X_i}= A_i X_i - C_i P_{X_i},  \hspace{0.2in} i=1,2,3.
		\label{Linear_Hamilton}
	\end{aligned}
\end{equation}
With the column vector $X=(X_1, X_2,X_3,P_{X_1}, P_{X_2}, P_{X_3})^T$, we can rewrite \eqref{Linear_Hamilton} in a matrix form,
\begin{equation}
	\dot X= MX + DX,
\end{equation}
where $M$ and $D$ represent the conservative  and dissipative parts of the dynamics, respectively,
\begin{equation}
	M=
	\begin{pmatrix}
		0 & 0 & 0 & 1 & 0 & 0\\
		0 & 0 & 0 & 0 & 1 & 0\\
		0 & 0 & 0 & 0 & 0 & 1\\
		A_1 & 0 & 0 & 0 & 0 & 0\\
		0 & A_2 & 0 & 0 & 0 & 0\\
		0 & 0 & A_3 & 0& 0 & 0\\
	\end{pmatrix},
	\hspace{0.1in}
	D=
	\begin{pmatrix}
		0 & 0 & 0 & 0& 0& 0\\
		0 & 0 & 0 & 0& 0& 0\\
		0 & 0 & 0 & 0& 0& 0\\
		0 & 0 & 0 & -C_1& 0& 0\\
		0 & 0 & 0 & 0& -C_2& 0\\
		0 & 0 & 0 & 0& 0& -C_3\\
	\end{pmatrix}.
\end{equation}
The corresponding quadratic Hamiltonian function for the linearized system is given by,
\begin{equation}
	\mathcal{H}=\tfrac{1}{2} \left(P_{X_1}^2 +P_{X_2}^2 + P_{X_2}^2 \right) - \tfrac{1}{2} \left(A_1 X_1^2 + A_2 X_2^2 + A_3 X_3^2\right).
	\label{quadratic_Hamiltonian}
\end{equation}

\section{Conservative system}
\label{linearized_conservative}
\subsection{Analytical solutions near the equilibria}
In this section, we analyze the linearized dynamics around the saddle in the conservative system with $C_i=0$ $\left(i=1,2,3 \right)$.  
The characteristic polynomial is,
\begin{equation}
	\begin{aligned}
		p(\beta)&=\beta^6 - \left(A_1 + A_2 + A_3 \right) \beta^4 + \left(A_1 A_2 + A_1 A_3 + A_2 A_3 \right) \beta^2 - A_1 A_2 A_3\\
		&= \left(\beta^2 - A_1 \right) \left(\beta^2 - A_2 \right) \left(\beta^2 - A_3 \right),
	\end{aligned}
\end{equation}
where $\beta$ denotes the eigenvalue. Notice that $A_1>0$, $A_2<0$, and $A_3<0$. In this case, the linearized system has one pair of real eigenvalues with opposite sign and two pairs of pure complex conjugate eigenvalues, denoted by  $\beta_{1,2}=\pm \lambda$, $\beta_{3,4}=\pm i \omega_2$, and $\beta_{5,6}=\pm i \omega_3$, where $\lambda=\sqrt{A_1}$, $\omega_2=\sqrt{-A_2}$, and $\omega_3=\sqrt{-A_3}$ are positive real constants. It shows that the equilibrium point is a saddle$\times$center$\times$center. After obtaining the eigenvalues, it is natural to obtain the corresponding eigenvectors which are denoted by $u_{\pm \lambda}$, $u_{\omega_2} \pm i v_{\omega_2}$, and $u_{\omega_3} \pm i v_{\omega_3}$, respectively. Introduce a linear change of variables given by,
\begin{equation}
	X= C Q,
	\label{change of variables}
\end{equation}
where $Q=\left(q_1,q_2,q_3,p_1,p_2,p_3 \right)^T$. The columns of the 6 $\times$ 6 matrix $C$ are given by the eigenvectors with the following form,
\begin{equation}
	C= \left(u_{\lambda}, u_{\omega_2}, u_{\omega_3},u_{-\lambda}, v_{\omega_2}, v_{\omega_3} \right)
	\label{general change of variable}
	=\begin{pmatrix}
		1 & 0 & 0 & -1 & 0 & 0\\
		0 & 1 & 0 & 0 & 0 & 0\\
		0 & 0 & 1 & 0 & 0 & 0\\
		\lambda & 0 & 0 & \lambda & 0 & 0 \\
		0 & 0 & 0 & 0 & \omega_2 & 0 \\
		0 & 0 & 0 & 0 & 0 & \omega_3 \\
	\end{pmatrix}.
\end{equation}
From \eqref{general change of variable}, we find the following relations,
\begin{equation}
	C^T J C=
	\begin{pmatrix}
		0  & B \\
		-B & 0
	\end{pmatrix}
	, \hspace{0.2in}
	B=
	\begin{pmatrix}
		2 \lambda & 0 & 0\\
		0 & \omega_2 & 0\\
		0 & 0 & \omega_3
	\end{pmatrix}.
\end{equation}
To obtain a symplectic change of variables \cite{MaRa1999,KoLoMaRo2011}, we need $C^T J C=J$, where $J$ is the $6 \times 6$
canoniical symplectic matrix defined by,
\begin{equation}
	J=\begin{pmatrix}
		0 & I_3\\
		-I_3& 0
	\end{pmatrix},
\end{equation}
and $I_3$ is the $3 \times 3$ identity matrix. Thus, we rescale the columns of $C$ by the factors $s_1=\sqrt{2 \lambda}$, $s_2=\sqrt{\omega_2}$, and $s_3=\sqrt{\omega_3}$ so that the final form of the symplectic matrix $C$ can be obtained as,
\begin{equation}
	C=
	\begin{pmatrix}
		\frac{1}{\sqrt{2 \lambda}} & 0 & 0 & -\frac{1}{\sqrt{2 \lambda}} & 0 & 0\\
		0 & \frac{1}{\sqrt{\omega_2}} & 0 & 0 & 0 & 0\\
		0 & 0 & \frac{1}{\sqrt{\omega_3}} & 0 & 0 & 0\\
		\sqrt{\frac{\lambda}{2}} & 0 & 0 & \sqrt{\frac{\lambda}{2}} & 0 & 0 \\
		0 & 0 & 0 & 0 & \sqrt{\omega_2} & 0 \\
		0 & 0 & 0 & 0 & 0 & \sqrt{\omega_3} \\
	\end{pmatrix},
	\label{simplectic transformation matrix}
\end{equation}
which casts the equations of motion into their symplectic eigenbasis normal form,
\begin{equation}
	\begin{aligned}
		& \dot q_1= \lambda q_1, \hspace{0.2in} && \dot p_1=-\lambda p_1,\\
		& \dot q_2= \omega_2 p_2, \hspace{0.2in} && \dot p_2 =-\omega_2 q_2,\\
		& \dot q_3= \omega_3 p_3, \hspace{0.2in} && \dot p_3 =-\omega_3 q_3,
		\label{normal_form_odes}
	\end{aligned}
\end{equation}
with quadratic Hamiltonian function given by,
\begin{equation}
	\mathcal{H}_2=\lambda q_1 p_1 + \tfrac{1}{2} \omega_2 \left(q_2^2 + p_2^2 \right) + \tfrac{1}{2} \omega_3 \left(q_3^2 + p_3^2 \right).
	\label{Hamiltonian normal form}
\end{equation}
The solutions of \eqref{normal_form_odes} can be conveniently written as,
\begin{equation}
	\begin{aligned}
		& q_1=q_1^0 e^{\lambda t}, \hspace{0.2in} p_1=p_1^0 e^{-\lambda t},\\
		& q_2 + i p_2= \left(q_2^0 + i p_2^0 \right) e^{-i \omega_2 t},\\
		& q_3 + i p_3= \left(q_3^0 + i p_3^0 \right) e^{-i \omega_3 t},
	\end{aligned}
	\label{sol_conser_eigen}
\end{equation}
where $Q_0=\left(q_1^0, q_2^0, q_3^0, p_1^0, p_2^0, p_3^0\right)^T$ are the initial conditions. Specifically under the Hamiltonian system \eqref{normal_form_odes} one can find the following independent constants of motion,
\begin{equation}
	f_1=q_1 p_1, \hspace{0.1in} f_2=q_2^2 + p_2^2, \hspace{0.1in} f_3=q_3^2 + p_3^2, 
	\label{constants of motion}
\end{equation}
one in each of the canonical projections $(q_i,p_i)$.

\subsection{Boundary of transit and non-transit orbits}

\begin{figure}[ht]
	\begin{center}
		\includegraphics[width=\textwidth]{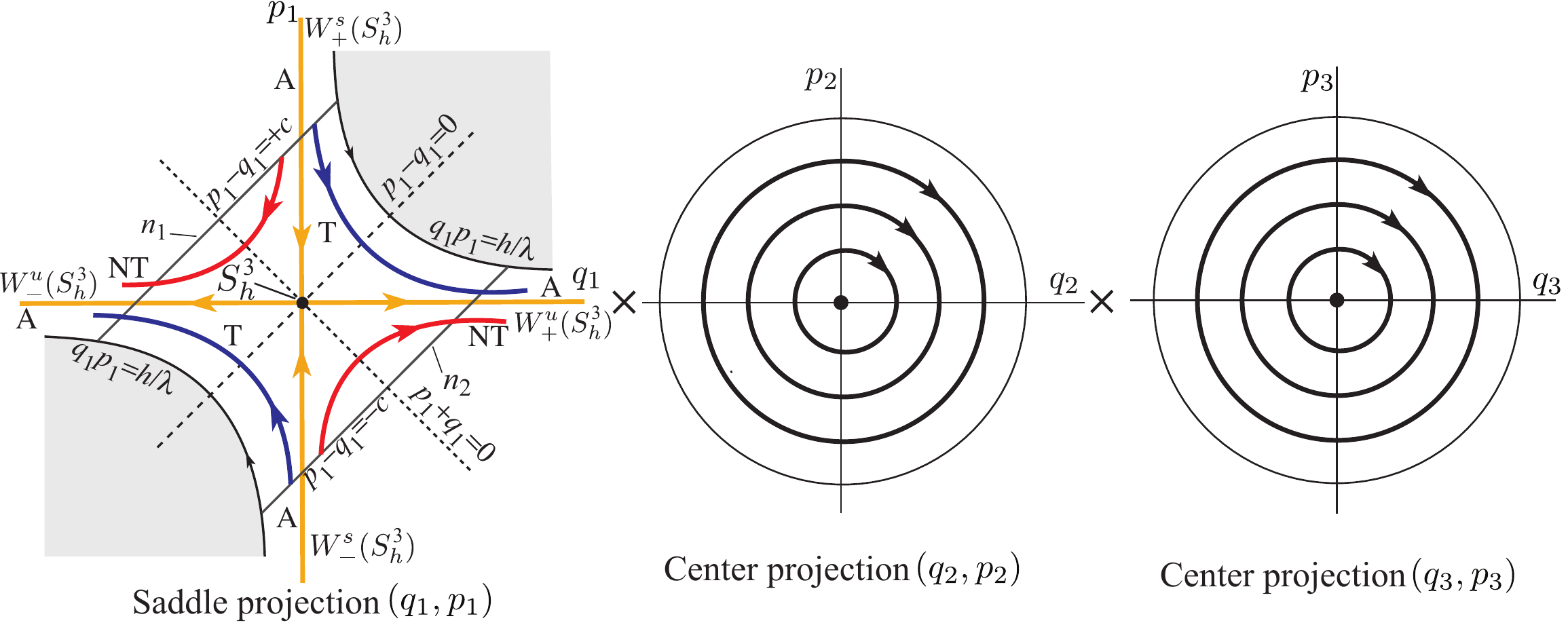}
	\end{center}
	\caption{\label{conservative_eigenspace_3DOF}{\footnotesize 
			The flow in the equilibrium region has the form
			saddle $\times$ center $\times$ center.
			On the left is shown the projection onto the $(p_1,q_1)$-plane, the saddle projection.
			Shown are the bounded orbits
			(black dot at the center), the asymptotic orbits (labeled A), two
			transit orbits (T) and two non-transit orbits (NT).}}
\end{figure}

\paragraph{The Linearized Phase Space} 
For positive $h$ and $c$,
the region $\mathcal{R}$, which is determined by
\begin{equation}
	\mathcal{H}_2=h, \quad  \mbox{and} \quad |p_1-q_1|\leq c,
\end{equation}
is homeomorphic to the product of a 4-sphere and an interval $I\subset \mathbb{R}$, 
$S^4\times I$;
namely, for each fixed value of $p_1 -q_1 $ in the interval $I=[-c,c]$,
we see that  the equation $\mathcal{H}_2=h$ determines a 4-sphere,
\begin{equation}
	\frac{\lambda }{4}(q_1 +p_1 )^2
	+ \frac{\omega_2 }{2}(q_2^2+p_2^2)
	+ \frac{\omega_3 }{2}(q_3^2+p_3^2)
	=h+\frac{\lambda }{4}(p_1 -q_1 )^2.
\end{equation}
The bounding 4-sphere of $\mathcal{R}$ for which $p_1 -q_1 = -c$ will
be called $n_1$, and  that where $p_1 -q_1 = c$,
$n_2$ (see Figure \ref{conservative_eigenspace_3DOF}).  

We call the set of points on each bounding 4-sphere where $p_1 + q_1 = 0$ the equator, and the sets where
$p_1 + q_1 > 0$ or $p_1 + q_1 < 0$ will be called the northern and 
southern hemispheres, respectively.

\paragraph{The Linear Flow in $\mathcal{R}$} 
To analyze the flow in
$\mathcal{R}$,  one considers the projections on the $(q_1, p_1)$-plane and $(q_2,p_2)\times (q_3,p_3)$-space, respectively.
In the  case of the $(q_1, p_1)$-plane, we see the standard picture of a saddle critical point.
In the case of the $(q_2,p_2)\times (q_3,p_3)$-space, we have a center manifold consisting of 
two uncoupled harmonic oscillators.
Figure \ref{conservative_eigenspace_3DOF} schematically illustrates the flow.
The coordinate axes of the
$(q_1,p_1)$-plane have been tilted by
$45^\circ$ and labeled $(p_1,q_1)$
in order to correspond to the
direction of the flow in later figures which adopt the NASA convention
that the larger primary is to the left of the smaller secondary.
With regard to the first projection we
see that $\mathcal{R}$ itself projects
to a set bounded on two sides by
the hyperbola
$q_1p_1 = h/\lambda $
(corresponding to $q_2^2+p_2^2=q_3^2+p_3^2=0$, see \eqref{Hamiltonian normal form}) and on two
other sides by the line segments
$p_1-q_1= \pm c$, which correspond to the bounding 4-spheres.

Since $q_1p_1$, from \eqref{constants of motion}, is an integral
of the equations in $\mathcal{R}$,
the projections of
orbits in the $(q_1,p_1)$-plane
move on the branches of the corresponding
hyperbolas $q_1p_1 =$ constant,
except in the case $q_1p_1=0$, where $q_1 =0$ or $p_1 =0$.
If $q_1p_1 >0$, the branches connect
the bounding line segments $p_1 -q_1 =\pm c$, that is, from $n_1$ to $n_2$ or vice versa. 
If $q_1p_1 <0$, the branches 
have both end points on the same segment, that is, 
from $n_1$ to $n_1$ or  $n_2$ to $n_2$.
A check of equation \eqref{normal_form_odes} 
shows that the orbits move
as indicated by the arrows in Figure \ref{conservative_eigenspace_3DOF}.

To interpret Figure \ref{conservative_eigenspace_3DOF} as a flow in $\mathcal{R}$,  notice
that each point in the $(q_1,p_1)$-plane projection 
corresponds to a 3-sphere $S^3$ in
$\mathcal{R}$ given by,
\begin{equation}
	\frac{\omega_2 }{2}(q_2^2+p_2^2)
	+ \frac{\omega_3 }{2}(q_3^2+p_3^2)
	=h-\lambda q_1p_1 .
\end{equation}
Of course, for points on the bounding  hyperbolic
segments ($q_1p_1 =h/\lambda $), the
3-sphere collapses to a point. Thus, the segments of the lines 
$p_1-q_1 =\pm c$  in the projection correspond to the 4-spheres
bounding $\mathcal{R}$.  This is because each corresponds to a
3-sphere crossed with an interval where the two end 3-spheres are
pinched to a point.

We distinguish nine classes of orbits grouped into the following four
categories:
\begin{enumerate}
	\item The point $q_1 =p_1 =0$ corresponds to an invariant
	3-sphere $S^3_h$ of {\bf bounded orbits} in
	$\mathcal{R}$, containing both {\bf periodic} and, more generally, {\bf quasi-periodic} orbits (see  \cite{levi2014classical}).
	This 3-sphere is given by,
	\begin{equation}\label{3-sphere}
		\frac{\omega_2 }{2}(q_2^2+p_2^2)
		+ \frac{\omega_3 }{2}(q_3^2+p_3^2)
		=h, \hspace{0.3in} q_1 =p_1 =0.
	\end{equation}
	It is an example of a 
	NHIM at a fixed energy $h$ (see \cite{Wiggins1994}).
	Roughly, this means that the stretching and contraction rates under
	the linearized dynamics transverse to the 3-sphere dominate those tangent
	to the 3-sphere.  This is clear for this example since the dynamics normal
	to the 3-sphere are described by the exponential contraction and expansion
	of the saddle point dynamics.  The 3-sphere acts as a ``big
	saddle point''.  
	See the black dot labeled $S^3_h$ at the center of the $(q_1,p_1)$-plane on the left side 
	of Figure
	\ref{conservative_eigenspace_3DOF}.
	
	\item The four half open segments on the axes, $q_1p_1 =0$, 
	correspond to four 
	cylinders of orbits asymptotic to this invariant 3-sphere 
	$S^3_h$ either as time
	increases ($p_1 =0$) or as time decreases ($q_1 =0$).  These are called {\bf
		asymptotic} orbits and they form the stable and the unstable manifolds of the 3-sphere of bounded orbits at this energy h,
	$S^3_h$.  The stable manifold branches, $W^s_{\pm}(S^3_h)$, are given by,
	\begin{equation}\label{stable_manifold}
		\frac{\omega_2 }{2}(q_2^2+p_2^2)
		+ \frac{\omega_3 }{2}(q_3^2+p_3^2)
		=h, \hspace{0.3in} q_1 =0.
	\end{equation}
	where 
	$W^s_+(S^3_h)$ is the branch with $p_1>0$ and
	$W^s_-(S^3_h)$ is the branch with $p_1<0$.
	The unstable manifold branches, $W^u_{\pm}(S^3_h)$, 
	are given by,
	\begin{equation}\label{unstable_manifold}
		\frac{\omega_2 }{2}(q_2^2+p_2^2)
		+ \frac{\omega_3 }{2}(q_3^2+p_3^2)
		=h, \hspace{0.3in} p_1 =0.
	\end{equation}
	where 
	$W^u_+(S^3_h)$ is the branch with $q_1>0$ and
	$W^u_-(S^3_h)$ is the branch with $q_1<0$.
	See the four orbits labeled A for asymptotic in Figure \ref{conservative_eigenspace_3DOF}.
	
	\item The hyperbola segments determined by
	$q_1p_1 ={\rm constant}>0$ correspond
	to two hyper-cylinders of orbits 
	which cross $\mathcal{R}$ from one bounding 4-sphere to the
	other, meeting both in the same hemisphere; the northern hemisphere
	if they go from
	$n_1$ $(p_1-q_1 =+c)$ to $n_2$ $(p_1-q_1 =-c)$, and the southern hemisphere
	in the other case. 
	Since these orbits transit from one potential well to another, or one side of the index-1 saddle point to the other, we call  them {\bf transit} orbits.  
	See the two orbits labeled T in Figure \ref{conservative_eigenspace_3DOF}.
	
	\item Finally the hyperbola segments determined by $q_1p_1 = {\rm
		constant}<0$ correspond to two hyper-cylinders of orbits in
	$\mathcal{R}$ each of which runs from one hemisphere to the other hemisphere
	on the same bounding 4-sphere.  
	Thus if $p_1 >0$, the 4-sphere is $n_1$ ($p_1-q_1 =+c$) and orbits run from 
	the northern hemisphere ($p_1 + q_1 >0$) 
	to the southern hemisphere ($p_1 +q_1 <0$) while the converse holds if $p_1 <0$, where the 
	4-sphere is $n_2$. Since these orbits return to the same potential well, we call them 
	{\bf non-transit} orbits.  See the two orbits labeled NT in Figure \ref{conservative_eigenspace_3DOF}.
\end{enumerate}

\paragraph{Invariant Manifolds as Separatrices}
The key observation here is that the asymptotic orbits form 4-dimensional stable
and unstable invariant manifold hyper-cylinders or ``tubes'' (with topology $S^3\times {\mathbb R}$), 
which are asymptotic to the invariant 3-sphere $S^3_h$ in a 5-dimensional energy surface  and 
which separate two distinct types of motion: transit orbits and non-transit orbits. 
The transit orbits, passing from one potential energy well to another, are those located inside the
4-dimensional manifold tube.  The non-transit orbits, which bounce back to
their potential energy well of origin, are those outside the tube.  

In fact, it can be shown that for a value of the Hamiltonian energy  just above that
of the index-1 saddle point, the nonlinear dynamics in the equilibrium region
${\mathcal R}$  is qualitatively the same as the linearized
picture that we have shown above.  

For example, 
the NHIM
for the nonlinear system which corresponds  to the 3-sphere
(\ref{3-sphere}) for the linearized system is given by,
\begin{equation}\label{NHIM_nonlinear}
	\mathcal{M}^3_h=\left\{(q,p)\mid \hspace{0.1in} 
	\tfrac{\omega_2 }{2}\left(q_2^2+p_2^2\right)+
	\tfrac{\omega_3 }{2}\left(q_3^2+p_3^2\right)+f(q_2,p_2,q_3,p_3)=h, \hspace{0.1in} 
	q_1=p_1=0\right\}
\end{equation}
where $f$ is at least of third-order.  Here, $(q_1,p_1,q_2,p_2,q_3,p_3)$
are normal form coordinates and are related to the linearized coordinates
via a near-identity transformation.

In a small neigborhood of the index-1 saddle equilibrium point,
since the nonlinear terms are much smaller than the linear terms,
the 3-sphere for the linear problem becomes a deformed sphere for the 
nonlinear problem.  
Moreover, since NHIMs persist under perturbation,
this deformed sphere $\mathcal{M}^3_h$ still has stable and unstable
manifolds which are given by,
\begin{equation}
	\begin{split}
		W^s_{\pm }(\mathcal{M}^3_h)&= \left\{(q,p)\mid \hspace{0.03in} 
		\tfrac{\omega_2 }{2}\left(q_2^2+p_2^2\right)+
		\tfrac{\omega_3 }{2}\left(q_3^2+p_3^2\right)+f(q_2,p_2,q_3,p_3)=h, 
		\hspace{0.03in} q_1=0 
		\right\}\\
		W^u_{\pm }(\mathcal{M}^3_h)&= \left\{(q,p)\mid \hspace{0.03in}
		\tfrac{\omega_2}{2}\left(q_2^2+p_2^2\right)+
		\tfrac{\omega_3}{2}\left(q_3^2+p_3^2\right)+f(q_2,p_2,q_3,p_3)=h, 
		\hspace{0.03in}
		p_1=0 \right\}.
	\end{split}
\end{equation}
Notice the similarity between the formulas above and those for the
linearized problem, \eqref{stable_manifold} and \eqref{unstable_manifold}.
See \cite{Hartman1964,JoMa1999,wiggins2001impenetrable} for details.  
This geometric insight will be used 
below to guide our numerical explorations in constructing transit orbits.

\subsection{Trajectories in the position space}
After considering the flow in the eigenspace, we now examine the appearance of the orbits in the position space. 
It should be mentioned that in position space, all possible motions are confined within the energy manifold, the boundary of which is a zero velocity or zero momentum surface \cite{Szebehely1967,KoLoMaRo2011} (corresponding to $P_{X_1}=P_{X_2}=P_{X_3}=0$), which can be obtained from \eqref{quadratic_Hamiltonian} as,
\begin{equation}
	A_1 X_1^2 + A_2 X_2^2 + A_3 X_3^2=-2h.
\end{equation}

Based on the solutions in \eqref{sol_conser_eigen} for the eigenspace in the conservative system, one can obtain the solutions in the phase space analytically as functions of time,
\begin{equation}
	\begin{aligned}
		& X_1=\frac{1}{\sqrt{2 \lambda}} q_1^0 e^{\lambda t} - \frac{1}{\sqrt{2 \lambda}} p_1^0 e^{-\lambda t}, \hspace{0.1in} && X_j=\frac{1}{\sqrt{\omega_i}} \left( q_j^0 \cos \omega_j t + p_j^0 \sin \omega_j t \right),\\
		& P_{X_1}=\sqrt{\frac{\lambda}{2}} q_1^0 e^{\lambda t} + \sqrt{\frac{\lambda}{2}} p_1^0 e^{-\lambda t}, \hspace{0.1in} && P_{X_j}= \sqrt{\omega_j} \left( -q_j^0 \sin \omega_j t + p_i^0 \cos \omega_j t \right).\\
	\end{aligned}
	\label{gener_sol_posit_space}
\end{equation}
Here and in the following, if not specifically pointed out, $i=2,3$.  

\begin{figure}[t]
	\begin{center}
		\includegraphics[width=\textwidth]{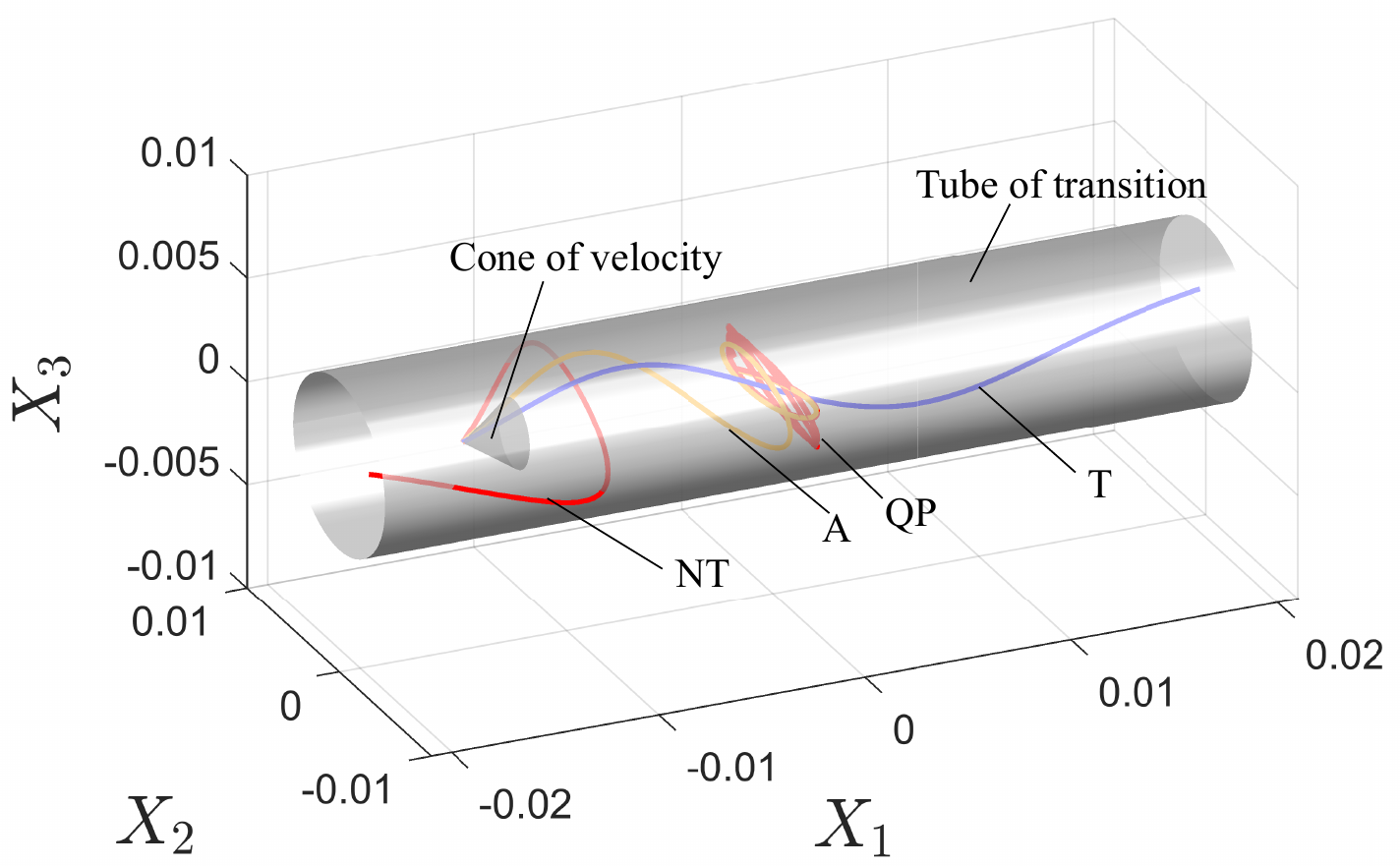}
	\end{center}
	\caption{{\footnotesize 
			Transition criteria and position space geometry in the conservative system with a fixed positive energy, $\mathcal{H}_2=h>0$. The tube of transition gives the possible initial positions of the transit orbits which means the transit orbits must start from a position within the tube. Moreover, for each position inside of the tube of transition, there exists a corresponding cone of velocity giving the directions of the velocity inside of which transit orbits occur. The transit orbit must have the initial velocity interior to the cone of velocity. Four types of trajectories are given in the figure: quasi-periodic orbit (QP), asymptotic orbit (A), transit orbit (T), and non-transit orbit (NT). Notice that the transit orbit, non-transit orbit, and asymptotic orbit all start from a position inside of the tube of transition. However, different initial velocity yields different type of orbits. The transit orbit and non-transit orbit have initial velocity inside and outside of the cone of velocity, while the asymptotic orbit have initial velocity on the boundary of the cone.
	}}
	\label{tube_of_transition}
\end{figure}

Inspecting the above general solutions and examining the limiting cases of the $X_1$ coordinate as $t$ goes to positive and negative infinity ($X_1$ acts as a `reaction coordinate' \cite{DeLi1994,collins2012isomerization}),
we can classify the same classes of orbits into the same four categories according to different combination of the signs of $p_1^0$ and $q_1^0$:
\begin{enumerate} 
	\item If $q_1^0=p_1^0=0$, the motion is on the center manifold (left two projections of Figure \ref{conservative_eigenspace_3DOF}), which could be periodic or quasi-periodic motion.

	\item Orbits with $q_1^0 p_1^0=0$ are asymptotic orbits. They are asymptotic to the periodic orbit. Asymptotic orbits with either $q_1^0$ or $p_1^0=0$ project into a cylinder, as shown in Figure \ref{tube_of_transition}, bounded by a surface defined by
	\begin{equation}
		\frac{\left(X_{2}^0 \right)^2}{ \left( \frac{\sqrt{2 h}}{\omega_2 }\right)^2 } + \frac{\left(X_{3}^0 \right)^2}{ \left( \frac{\sqrt{2 h}}{\omega_3 }\right)^2 }=1.
	\end{equation}
	
	\item Orbits with $q_1^0 p_1^0>0$ are transit orbits since when $q_1^0>0$ and $p_1^0>0$, the orbits cross the equilibrium region from $- \infty$ (bottom) to $+ \infty$ (top) or vice versa.

	\item Orbits with $q_1^0 p_1^0<0$ are non-transit orbits. When $q_1^0<0$ and $p_1^0>0$, the orbits starting from $-\infty$ approach to $-\infty$. When $q_1^0>0$ and $p_1^0<0$, the orbits starting from $+\infty$ approach to $+\infty$.
\end{enumerate}

\paragraph{Tube of transition and cone of velocity}

Defining the initial conditions in the phase space as $X_0= \left(X_1^0, X_2^0, X_3^0, P_{X_1}^0, P_{X_2}^0, P_{X_3}^0 \right)$, one can connect $X_0$ and $Q_0$ by the symplectic change of variables in \eqref{change of variables}, i.e., $X_0=C Q_0$.
As discussed in the previous section, the stable manifold tubes consisting of stable asymptotic orbits separate the transit orbits and non-transit orbits. For the stable asymptotic orbits, we have $q_1^0=0$. Thus, we have the following relations for  stable asymptotic orbits,
\begin{equation}
	q_1^0=0, \hspace{0.1in}q_2^0=\sqrt{\omega_2} X_2^0, \hspace{0.1in} q_3^0=\sqrt{\omega_3} X_3^0, \hspace{0.1in} p_1^0=-\sqrt{2 \lambda} X_1^0, \hspace{0.1in} p_2^0=\frac{P_{X_2}^0}{\sqrt{\omega_2}}, \hspace{0.1in} p_3^0=\frac{P_{X_3}^0}{\sqrt{\omega_3}}.
	\label{ini_cond_cons_eigen}
\end{equation}
The substitution of the relations in \eqref{ini_cond_cons_eigen} into the Hamiltonian normal form in \eqref{Hamiltonian normal form} gives the initial conditions in the phase space for the stable asymptotic orbits,
\begin{equation}
	\frac{\left(X_{2}^0 \right)^2}{ a_{X_2}^2} + \frac{\left(X_{3}^0 \right)^2}{ a_{X_3}^2 }+\frac{\left(P_{X_2}^0 \right)^2}{ \left( \sqrt{2 h}\right)^2 } + \frac{\left(P_{X_3}^0 \right)^2}{ \left( \sqrt{2 h}\right)^2 }  =1, \hspace{0.1in} \text{and } P_{X_1}^0=-\lambda X_1^0,
	\label{transition tube in Cartesian}
\end{equation}
where $a_{X_2}=\sqrt{2h}/\omega_2$, $a_{X_3}=\sqrt{2h}/\omega_3$.
Notice that the six phase space variables are not all independent. The four variables
$\left(X_{2},X_{3},P_{X_2},P_{X_3}\right)$, according to the first equation in \eqref{transition tube in Cartesian}, topologically describe a 3-sphere.
The two variables
$\left(X_{1},P_{X_1}\right)$ describe a line. 
We refer to the direct product of a 3-sphere and a line segment as a hyper-cylinder (or colloquially as a tube).
It is the transition boundary separating the transit and non-transit orbits starting at an initial energy $h$,
which we will call $\partial \mathcal{T}_h$.
As it is a tube, we refer to this 4-dimensional object $\partial \mathcal{T}_h$ as the {\bf transition tube} of energy $h$.
Although the analytical form of the transition criteria have been derived 
in \eqref{transition tube in Cartesian}, it is impossible to plot the geometric structures in the 
6-dimensional phase space. 
In the following, we focus on discussing the transition criteria and geometry in the three-dimensional position space, as it is easier to interpret.

Since the last two terms related to the momenta at the left side of \eqref{transition tube in Cartesian} should be non-negative, we have that the projection of  $\partial \mathcal{T}_h$ onto configuration space satisfies,
\begin{equation}
	(X_2^0/a_{X_2})^2 + (X_3^0/a_{X_3})^2 \leq 1.
\end{equation}
The form in the position space appears as a solid tube, in the  sense of disc $D^2 \times \mathbb{R}$. (where  $D^2$ is the four-dimensional closed disc) with boundary of a cylinder $S^1 \times \mathbb{R}$.
Here we refer to it as the {\bf tube of transition}, and it is the 3-degree-of-freedom counterpart 
of the strips of transition described by Conley \cite{Conley1968} and Koon et al.\ \cite{KoLoMaRo2011} for a 2-degree-of-freedom Hamiltonian system. 
The tube of transition confines the existence of transit orbits of the given energy $h$; transit orbits can only start with positions within the tube. 
Outside of the tube, and at the same energy $h$, the situation is simple. 
All trajectories starting there are non-transit orbits. 
It means the sign of $q_1^0 p_1^0$  in the eigenspace variables is always negative, independent of the direction of the velocity. For example, the signs in each component of the equilibrium region $\mathcal{R}$ complementary to the tube can be determined by limiting behavior of $X_1$ for positive and negative infinite time. For example, in the left component the non-transit orbits stay on the left side for $t \rightarrow \pm \infty$, indicating $p_1^0>0$ and $q_1^0<0$. Similarly, in the right component we have $p_1^0<0$ and $q_1^0>0$.

For a specific position $(X_1^0, X_2^0, X_3^0)$ inside the tube, the situation is more complicated since the initial position within the tube is only a necessary condition for a transit orbit and the signs of $q_1^0$ and $p_1^0$ are dependent on direction of the velocity. 
In the following we aim to derive another criteria in addition to the tube of transition for the transit orbits with the position inside the tube. 
We introduce $R$, $\theta$, and $\phi$ to rewrite the three momenta as in spherical coordinates,
\begin{equation}
	P_{X_1}^0=R \sin \theta, \hspace{0.2in} P_{X_2}= R \cos \theta \sin \phi, \hspace{0.2in} P_{X_3}=R \cos \theta \cos \phi,
	\label{spherical_coordinate}
\end{equation}
where $R$, the momentum magnitude, is given by,
\begin{equation}
	R=2h -  \left[ A_1 \left( X_1^0 \right)^2  + A_2 \left( X_2^0 \right)^2 + A_3 \left( X_3^0 \right)^2\right].
\end{equation}
From the relation $P_{X_1}^0=-\lambda X_1^0=R \sin \theta$, one can obtain the conditions on the angles,
\begin{equation}
	\theta= \arcsin \left(-\frac{\lambda X_1^0}{R} \right), \hspace{0.2in} \phi \in [0, 2 \pi].
	\label{cone_conservative}
\end{equation} 
As a result, the above relations together define a cone at the position $(X_1^0, X_2^0, X_3^0)$ within the tube of transition. Here we refer to it as the {\bf cone of velocity}, in analogy with the wedge of velocity discussed by Conley \cite{Conley1968} (see also \cite{zhong2020geometry}).
Notice the cone only exists inside of the tube of transition to ensure $R>0$. 

Figure \ref{tube_of_transition} gives the flow of the conservative system projected onto the position space. A tube of transition of a given energy is shown; that is, the projection of the position-based necessary condition. The orbits outside the tube are all non-transit orbits. Inside the tube, the cone of velocity acts as a secondary condition to guarantee the transition. 
It is the separatrix of transit and non-transit orbits at each point in the position-based  tube of transition. 
The orbits with velocity inside of the cone are  transit orbits, while those with velocity outside of the cone are non-transit orbits.
Orbits with velocity on the boundary of the cone are the stable asymptotic orbits. 
In Figure \ref{tube_of_transition} , the four types of orbits mentioned above are given.

One can notice from the infinite length of the tube along the $X_1$ direction that the transit orbits can start from a position infinitely far from the equilibrium point due to energy conservation. However, in a practice, this is not observed since  energy dissipation cannot be avoided in the real world. In the next section, we  drop the ideal case of energy conservation and study a more practical case where energy dissipation is included and consider the geometry and criteria for transition in the dissipative system.

\section{Dissipative system}
\label{linearized_dissipative}
\subsection{Analytical solutions near the equilibria}
In this section, we focus on the linearized dynamics around the index-1 saddle in the dissipative systems. Using the same change of variables in \eqref{change of variables}, the equations of motion in the dissipative system can be written as,
\begin{subequations}	
	\begin{align}
		&\begin{cases}
			\dot q_1 = \left(\lambda - \frac{ C_1}{2}   \right) q_1 - \frac{ C_1}{2} p_1,\\
			\dot p_1 = - \frac{ C_1}{2} q_1 + \left(- \lambda -\frac{ C_1 }{2}  \right) p_1,	
		\end{cases}\\
		&\begin{cases}
			\dot q_j = \omega_j p_j,\\
			\dot p_j = - \omega_j q_j - C_j p_j.
		\end{cases}		
	\end{align}
	\label{ODEs_diss_symplect}	
\end{subequations}
Note that the dynamics on the $(q_1,p_1)$, $(q_2,p_2)$, and $(q_3,p_3)$ planes are 
uncoupled, which makes them easily solved analytically. The characteristic polynomial for each plane is given by,
\begin{equation}
	\begin{aligned}
		&p_1 \left(\beta \right)=\beta^2 + C_1 \beta - \lambda^2,\\
		& p_j \left(\beta \right)=\beta^2 + C_j \beta + \omega_j^2,
	\end{aligned}
\end{equation}
with the following eigenvalues,
\begin{equation}
	\begin{aligned}
		& \beta_{1,2}=\frac{-C_1\pm \sqrt{C_1^2 + 4 \lambda^2}}{2},\\
		& \beta_{2j-1,2j}= - \delta_j \pm i \omega_{d_j},\\
	\end{aligned}
\end{equation}
where,
\begin{equation}
	\begin{aligned}
		\delta_j =C_j/2, \hspace{0.1in} \omega_{d_j}=\omega_j \sqrt{1 - \xi_{d_j}^2}, \hspace{0.1in} \xi_{d_j}=\delta_j / \omega_j.
	\end{aligned}
\end{equation}
Thus, the general solutions are given by, 
\begin{subequations}	
	\begin{align}
		&\begin{cases}
			q_1= k_1 e^{\beta_1 t} + k_2 e^{\beta_2 t},\\
			p_1= k_3 e^{\beta_1 t} + k_4 e^{\beta_2 t},
		\end{cases}\\
		&\begin{cases}
			q_j= k_{2j+1}  e^{- \delta_j t} \cos{\omega_{d_j} t} + k_{2j+2} e^{- \delta_j t} \sin{\omega_{d_j} t},\\
			p_j= \frac{k_{2j+1} }{\omega_j} e^{- \delta_j t} \left(-\delta_j \cos{\omega_{d_j} t} - \omega_{d_j} \sin{\omega_{d_j} t} \right) +\frac{k_{2j+2} }{\omega_j}  e^{- \delta_j t} \left(\omega_{d_j} \cos{\omega_{d_j} t - \delta_j \sin{\omega_{d_j} t}} \right).
		\end{cases}
	\end{align}
	\label{sol_diss_symplect}
\end{subequations}
where,
\begin{equation*}
	\begin{aligned}
		&k_1 = \frac{q_1^0 \left(2 \lambda + \sqrt{C_1^2 + 4 \lambda^2} \right)-C_1 p_1^0 }{2\sqrt{C_1^2 + 4 \lambda^2}},  && k_2 =\frac{q_1^0 \left(-2 \lambda + \sqrt{C_1^2 + 4 \lambda^2} \right)+C_1 p_1^0 }{2\sqrt{C_1^2 + 4 \lambda^2}},\\
		&k_3 = \frac{p_1^0 \left(-2 \lambda + \sqrt{C_1^2 + 4 \lambda^2} \right)-C_1 q_1^0 }{2\sqrt{C_1^2 + 4 \lambda^2}},  && k_4 = \frac{p_1^0 \left(2 \lambda + \sqrt{C_1^2 + 4 \lambda^2} \right)+C_1 q_1^0 }{2\sqrt{C_1^2 + 4 \lambda^2}},\\
		&k_{2j+1}=q^0_j ,  &&k_{2j+2}=\frac{p^0_j \omega_j + q^0_j \delta_j}{\omega_{d_j}}. 
	\end{aligned}
\end{equation*}

Taking the total derivative of the Hamiltonian function in \eqref{Hamiltonian normal form} and applying the equations of motion in \eqref{ODEs_diss_symplect}, we have,
\begin{equation}
	\dot{\mathcal{H}}_2= - \tfrac{1}{2} C_1 \lambda (q_1 + p_1)^2 -  C_2 \omega_2 p_2^2 -  C_3 \omega_3 p_3^2 \leq 0,
\end{equation}
which means the Hamiltonian or the total energy is always decreasing (more precisely, non-increasing) due to the presence of damping. 

\subsection{Boundary of transit and non-transit orbits}

\begin{figure}[t]
	\begin{center}
		\includegraphics[width=\textwidth]{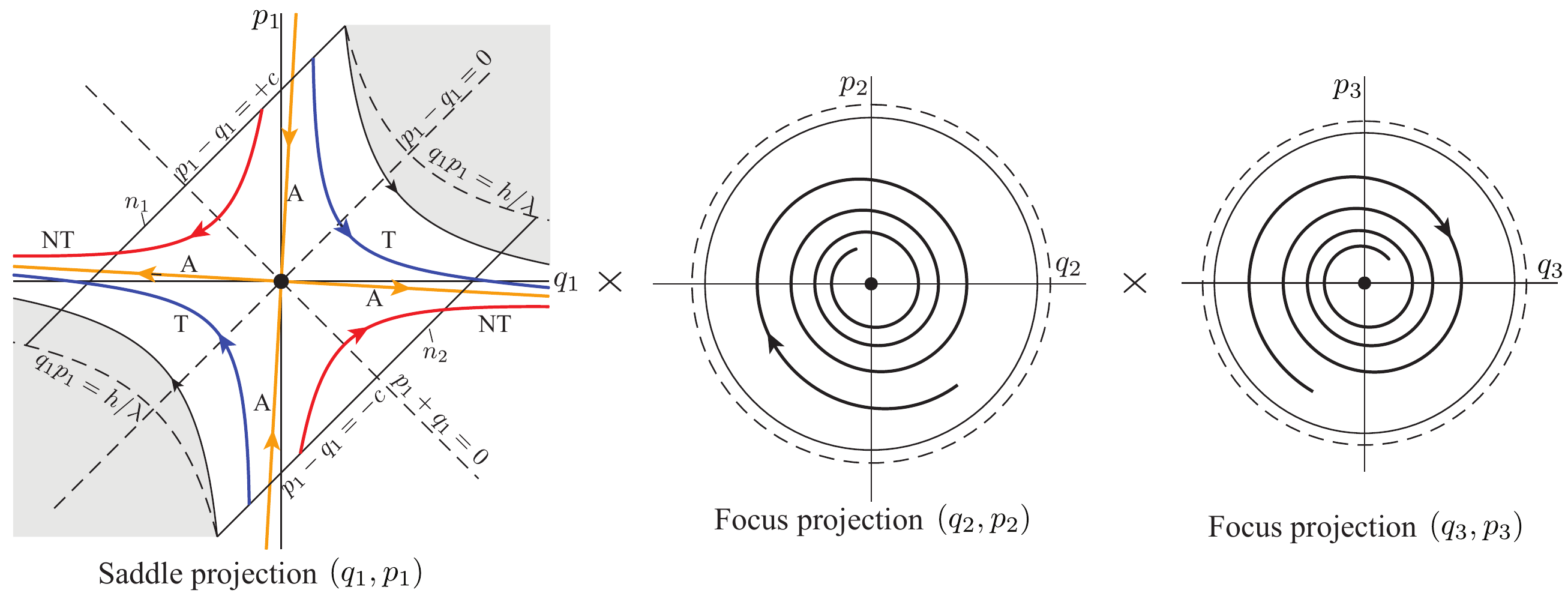}
	\end{center}
	\caption{{\footnotesize 
			The flow in the equilibrium region of the dissipative system has the form saddle$\times$focus$\times$focus. The black dot in the saddle projection corresponds to the focus-type asymptotic orbits whose motions are only on the two focus projections. The four half open segments, defined by $q_1=C_1 p_1/(2 \lambda \pm \sqrt{C_1^2 + 4 \lambda^2})$, are the saddle-type asymptotic orbits (A). The saddle-type asymptotic orbits in the dissipative system are titled clockwise compared to the asymptotic orbits in the conservative system. The stable and unstable asymptotic orbits are the stable and unstable invariant manifolds of the equilibrium point, respectively. Especially, the stable invariant manifold is the separatrix between transit orbits (T) and non-transit orbits (NT). The trajectories on the boundary of the shaded region in the saddle projection is the fastest trajectories with initial conditions on the bounding sphere in the dissipative system, while the dashed trajectories are the fastest trajectories in the conservative system. The dash circles and solid circles on the two focus projections are the boundary of the initial conditions for the transit orbits in the conservative and dissipative systems, respectively. 
	}}
	\label{eigenspace_dissipative}
\end{figure}

\paragraph{The linear flow in $\mathcal{R}$} 
Analogous to the discussion in the conservative system, the same equilibrium region $\mathcal{R}$ is selected to show the projections in the $(q_1,p_1)$-plane, $(q_2,p_3)$-plane, and $(q_3,p_3)$-plane. We note from the solution in \eqref{sol_diss_symplect} that in contrast to the saddle $\times$ center $\times$ center projections in the conservative system, the dissipative system presents saddle $\times$ focus $\times$ focus projections. The two focus projections are damped oscillators with frequencies given by $\omega_{d_j}=\omega_j (1-\xi_{d_j}^2)^{1/2}$, $j=2,3$. Observing the limiting cases of $X_1$ as $t \rightarrow \pm  \infty$, we can classify the orbits into the following four categories:
\begin{enumerate}
	\item The origin in the saddle projection, e.g., $(q_1,p_1)=(0,0)$, corresponds to a {\bf focus-type asymptotic} orbit which only has motion in the two focus projections, e.g., $(q_2,p_2)$-plane and $(q_3,p_3)$-plane. See the dot at the origin of the $(q_1,p_1)$-plane in Figure \ref{eigenspace_dissipative}. Due to the effect of damping, the periodic orbit in the conservative system does not exist. 
	
	\item The four half open segments on the lines defined by $q_1=C_1p_1/(2 \lambda \pm \sqrt{C_1^2 + 4 \lambda^2})$ are {\bf saddle-type asymptotic} orbits. See the four orbits labeled A in Figure \ref{eigenspace_dissipative}. Compared to the asymptotic orbits in the conservative system, the saddle-type asymptotic orbits tilted counterclockwise with a certain angle which is dependent on the magnitude of the damping. Moreover, the saddle-type asymptotic orbits are asymptotic to the equilibrium point which is different from those in the conservative cases asymptotic to the periodic orbits. 
	
	\item The segments which cross $\mathcal{R}$ from one bounding line to another, i.e., from $p_1-q_1=+c$ to $p_1-q_1=-c$ in the northern hemisphere, and vice versa in the southern hemisphere, correspond to {\bf transit orbits}. See the orbits labeled T of Figure \ref{eigenspace_dissipative}. Notice that segments giving the initial conditions of transit orbits in the dissipative system are shorter than those in the conservative system which means the damping reduces the amount of transit orbits.
	
	\item Finally the segments which run from one hemisphere to the other hemisphere on the same boundary, namely which stat from $p_1-q_1=\pm c$ and return the the same bounding line, correspond to {\bf non-transit} orbits. See the two orbits labeled NT of Figure \ref{eigenspace_dissipative}. In the dissipative system, the segments for the initial conditions of the non-transit orbits are longer than those in the conservative system. 
	
\end{enumerate}

\paragraph{Stable invariant manifold of the saddle point as separatrices} 
We find that the asymptotic orbits are the stable and unstable invariant manifolds of the equilibrium point itself (rather than a NHIM restricted to an energy surface). 
The stable invariant manifold of a given energy appears as an ellipsoid and it separates the initial conditions of orbits with distinct types of motion: transit orbits and non-transit orbits. Transit orbits passing from one potential well to another must have initial conditions inside of the ellipsoid, while non-transit orbits staying within the potential well where they come from have initial conditions outside of the ellipsoid.

\subsection{Trajectories in the position space}
After analyzing the trajectories in the eigenspace, we pay attention to the situation in the position space. 
From the analytical solutions in \eqref{sol_diss_symplect} for the eigenspace and the change of 
variables in \eqref{change of variables}, we can obtain the general solutions in the position space as,
\begin{equation}
	\begin{aligned}
		X_1 &= \frac{1}{\sqrt{2 \lambda}} \left(\bar k_1 e^{\beta_1 t} + \bar k_2 e^{\beta_2 t} \right),\\
		X_j & = \frac{1}{\sqrt{\omega_j}} e^{-\delta_j t} \left(k_{2j+1} \cos \omega_{d_2} t + k_{2j+2} \sin \omega_{d_j} t \right) ,\hspace{0.2in} j=2,3.
	\end{aligned}
\end{equation}
Here $\bar k_1=k_1 - k_3$ and $\bar k_2 =k_4 -k_2$. 

Analogous to the classification of the orbits in the conservative system, we can also classify the orbits 
according to the following four categories according to the limiting behavior of $X_1$ when 
$t$ approaches  positive and negative infinity. The four categories of orbits are,
\begin{enumerate}
	\item Orbits with $\bar k_1 =\bar k_2 =0$ are focus-type  asymptotic orbits. 
	\item Orbits with $\bar k_1 \bar k_2 =0$ are saddle-type  asymptotic orbits. 
	\item Orbits with $\bar k_1 \bar k_2 >0$ are transit orbits. 
	\item  Orbits with $\bar k_1 \bar k_2 <0$ are non-transit orbits. 
\end{enumerate}

Although we have recognized four different categories of orbits, the transition criteria and the geometry that governs the transition are still not clear. 
As discussed above, the transition boundary is given by the stable invariant manifold of the index-1 saddle. 
Thus, to get the transition boundary, we only need to compute the initial conditions of a given energy we are interested in for the stable asymptotic orbits. 
According to the above discussion, such initial conditions are only 
determined by the saddle projection $(q_1,p_1)$ given by,
\begin{equation}
	q_{1}^0=k_p p_{1}^0,
	\label{init_stab_invar_dissip}
\end{equation}
where $k_p=c_1/ \left(2 \lambda + \sqrt{c_1^2 + 4 \lambda^2} \right)$.
Thus, submitting the relation $Q_0=C^{-1} X_0$ for the initial conditions in the eigenspace into the Hamiltonian normal form in 
\eqref{Hamiltonian normal form}, and applying the relation in \eqref{init_stab_invar_dissip}, we can rewrite the Hamiltonian function in terms of the initial conditions in the phase space as,
\begin{equation}
	\frac{\left( X_1^0 \right)^2}{ a_{X_1}^2} + \frac{\left(X_{2}^0 \right)^2}{ a_{X_2}^2} + \frac{\left(X_{3}^0 \right)^2}{a_{X_3}^2}+ \frac{\left(P_{X_2}^0 \right)^2}{ \left( \sqrt{2 h}\right)^2 } + \frac{\left(P_{X_3}^0 \right)^2}{ \left( \sqrt{2 h}\right)^2 }=1,
	\label{transition_ellipsoid}
\end{equation}
where $a_{X_1} = \sqrt{\frac{h}{2 k_p}} \frac{1-k_p}{\lambda}$, and,
\begin{equation}
	P_{X_1}^0=\frac{k_p+1}{k_p-1} \lambda X_1^0.
	\label{X10_PX10}
\end{equation}
The form here is topologically a 4-sphere, geometrically a 4-dimensional ellipsoid in the 6-dimensional phase space.

Due to the non-negativity of the last two terms in the left hand side of \eqref{transition_ellipsoid} for the generalized momenta, we have following relation in the position space,
\begin{equation}
	\frac{\left( X_1^0 \right)^2}{ a_{X_1}^2} + \frac{\left(X_{2}^0 \right)^2}{ a_{X_2}^2} + \frac{\left(X_{3}^0 \right)^2}{a_{X_3}^2} \leq 1.
\end{equation}
The above relation defines a solid ellipsoid. It is referred to the ellipsoid of transition. 
All the transit orbits should have initial positions inside of the ellipsoid of transition. 
However, it only constrains the possible existence of the initial position for the asymptotic orbits. 
We also need to constrain the momenta or velocity. 
For an initial position inside the ellipsoid, from the definitions of $R$, $\theta$, and $\phi$ in \eqref{spherical_coordinate}, 
and relation in \eqref{X10_PX10}, we can obtain the following relation for the momenta,
\begin{equation}
	\theta = \arcsin \left( \frac{\lambda X_1^0 (k_p+1)}{R(k_p -1)}\right), \hspace{0.2in} \phi \in [0 , 2\pi].
	\label{cone_dissipative}
\end{equation}
Thus, at  position $(X_1^0,X_2^0,X_3^0)$ inside of the ellipsoid of transition, \eqref{cone_dissipative} 
defines a cone to constrain the velocity of the asymptotic orbits. It is the cone of velocity for the dissipative system. Comparing the cone of velocity for the conservative system in \eqref{cone_conservative} and the one for the dissipative system in \eqref{cone_dissipative}, we can notice that the energy dissipation or the damping decreases the size of the cone at a given specific position inside of the ellipsoid, corresponding to an decreasing number of transit orbits as a fraction of all possible orbits.

\begin{figure}[t]
	\begin{center}
		\includegraphics[width=\textwidth]{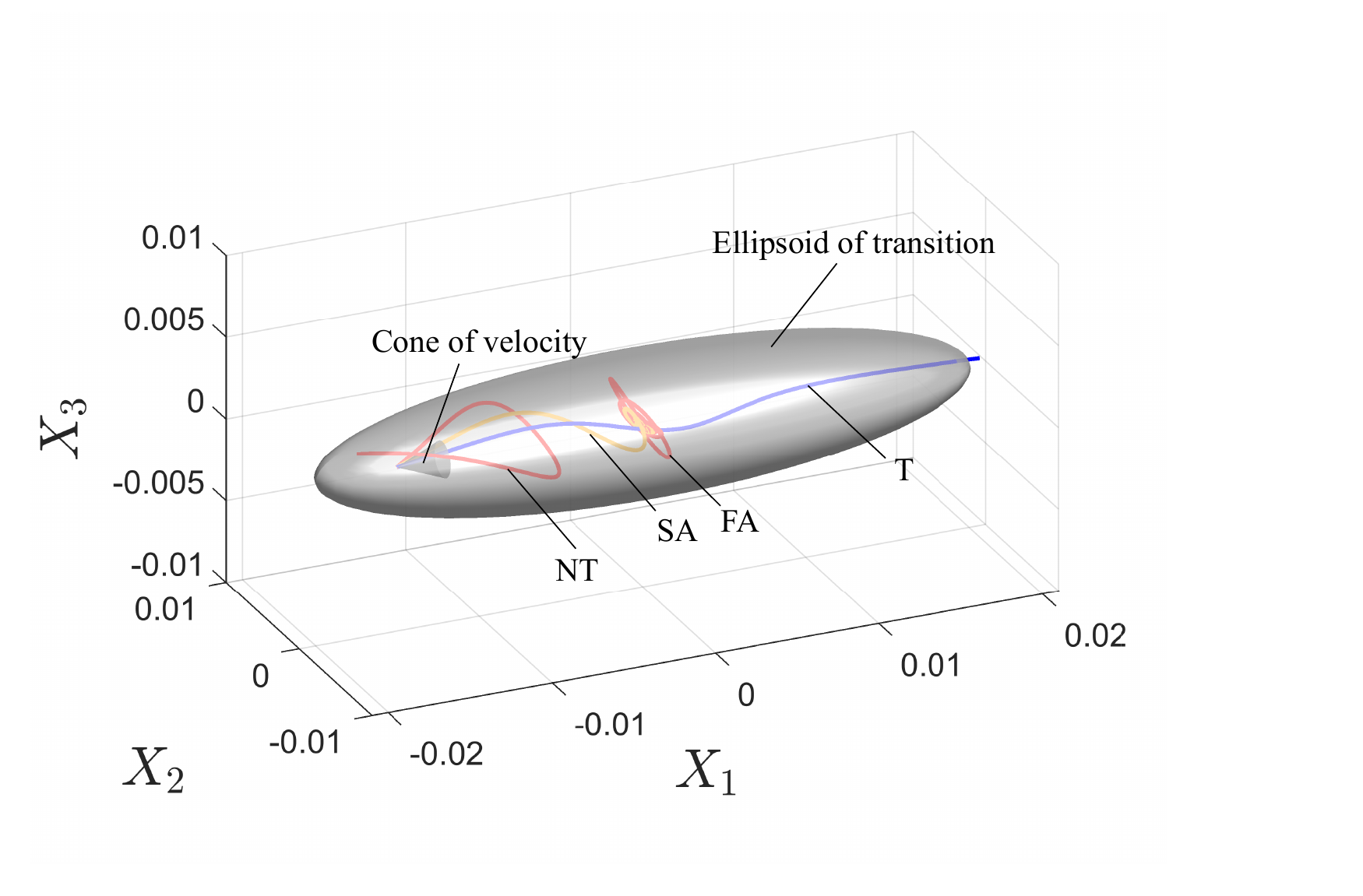}
	\end{center}
	\caption{{\footnotesize 
			Ellipsoid of transition and cone of velocity defining the transition criteria in the dissipative system. Transit orbits must start from a position inside of the ellipsoid of transition and their initial velocity must be interior to the cone of velocity. If either criterion is not satisfied, the trajectories can not escape from the potential well. In order to show the transition criteria, different types of orbits are given. Starting from a position inside of the ellipsoid of transition, the orbit with velocity interior to cone is transit orbit (T); the orbit with velocity outside of the cone is non-transit orbit (NT); The orbit with velocity on the boundary of the cone is saddle-type asymptotic orbit (SA). Moreover, a focus-type asymptotic orbit (FA) moving on the plane at $X_1$ is also shown. 
	}}
	\label{ellipsoid_of_transition}
\end{figure}

Figure \ref{ellipsoid_of_transition} gives the projections onto the position space around the equilibrium region. In contrast to the tube of transition that restricts the existence of the transit orbits in the conservative system, 
the geometric boundary for the possible existence of transit orbits in the dissipative system becomes an ellipsoid of transition. 
We  conclude that the damping `closes' the cylinder in the conservative system to form an ellipsoid in the dissipative system so that the farthest position for transit orbits are the end points of the ellipsoid. 
This means the region outside of the ellipsoid only allows  non-transit orbits. 
However, it does not mean the orbits that start from the region inside of the ellipsoid can definitely escape to the other potential well since it only confines the possible position of transit orbits. 
A sufficient condition is necessary, given by the cone of velocity. 
For each position inside of the ellipsoid, there exists a cone of velocity giving the right direction of transit orbits. 
The orbits with velocity interior to the cone are transit orbits, while those with velocity outside the cone are non-transit orbits. The boundary of cone gives the velocity for the asymptotic orbits. At different position, the size of the cone is different. Especially, on the surface of the ellipsoid of transition, only one asymptotic orbits exists. Moreover, at the same point inside of the ellipsoid, the size of the cone of velocity for the dissipative system is smaller than that for the conservative system. Figure \ref{ellipsoid_of_transition} also presents the ellipsoid of transition. At the same time, a cone of velocity at a specific position is also plotted. In order to show how the combination of the ellipsoid of transition and cone of velocity together defines the geometry and criteria of transition, four types of orbits are shown. We can notice that each orbit follows the established transition criteria.

\section{Discussions and remarks}
\label{Discussion}
In the above sections, we have discussed the geometry of escape in both the eigenspace and the position space. We found that in the position space, the starting position of a transit orbit that escape from one potential well to another must be inside of the tube and the ellipsoid in the conservative and dissipative systems, respectively. Apart from the necessary condition for the initial position, an sufficient condition for the transit orbits is required for the velocity to ensure the transition. At each point inside the position space structures, there exists a cone that gives the `right' direction approaching to the neck region. All transit orbits should start with an initial velocity interior to the cone. Thus, the tube of transition in the conservative system and the ellipsoid of transition in the dissipative system associated with the respective cone of velocity together give the transition criteria. To have a more intuitive interpretation, we recall the position space  geometry that governs the transition in the two degree of freedom systems \cite{zhong2018tube,zhong2020geometry,zhong2021global}. In such systems, the strip in the conservative system and the ellipse in the dssipative system combined with wedges of velocity can determine the initial conditions of a prescribed energy for the transit orbits. Here we can correspond the strip and ellipse to the tube and ellipsoid, and correspond the wedge of velocity to the cone of velocity, in the two degree of freedom systems and three degree of freedom systems.

As for the 6-dimensional phase space of the three degree of freedom system, the transition boundary of initial conditions of initial energy $h$, $\partial \mathcal{T}_h$, is a 4-dimensional object for both the conservative and dissipative systems. However, for the conservative case, it is a hyper-cylinder of open topology $S^3 \times \mathbb{R}$, whereas for the dissipative case, is a hyper-ellipsoid of compact topology $S^4$. We refer to them as the transition tube and transition ellipsoid, respectively. 
It was demonstrated that the transition tube and transition ellipsoid are the stable invariant manifolds of 
a set of bounded orbits (center manifold of the index-1 saddle of energy $h$) and of the index-1 saddle point itself, respectively. 
Since it is not possible to plot  manifolds with dimension higher than three, here we leave the hyper-cylinder and hyper-ellipsoid in the form of mathematical formulae given in \eqref{transition tube in Cartesian} and \eqref{transition_ellipsoid}, respectively. 
The transition boundary defines the initial conditions for  transit orbits of a certain  initial energy $h$. The interior region of the hyper phase space structures gives the initial conditions for the transit orbits, while the exterior region gives the initial conditions for the non-transit orbits. Finally, it should be pointed out that we need to distinguish the difference of the tube and ellipsoid in the position space and the phase space. Although they are tubes and ellipsoid, their dimensions and functionalities are physically and mathematically different. The ones in the position space only restrict the positions of the transit orbits which need to be combined with the cone of velocity to determine the initial conditions together. However, the dynamical structures in the phase space can individually determine the initial conditions.

When extending the linearized case to the nonlinear system, we can grow the local invariant manifold of the linearized system to the global invariant manifold. For the construction of the global invariant manifold, plenty computational methods have been proposed \cite{dellnitz1997subdivision,madrid2009distinguished,anderson2017isolating,zhong2021global,krauskopf2003computing,krauskopf2005survey,krauskopf2007numerical}. Among those algorithms, the idea of computing the global invariant manifold as a solution family of a suitable boundary-value problem (BVP) \cite{zhong2021global,krauskopf2007numerical} is a good candidate. During the application of continuation to obtain the solution families, the selection of the additional boundary condition apart from the ones approximated from the local invariant manifold is flexible, such as the time, arclength, or endpoint of the trajectories \cite{krauskopf2007numerical}. When implementing the BVP approach to compute higher dimensional invariant manifolds, we are faced with multi-parameter continuation which is complicated to be accomplished. Reference \cite{zhong2021global} proposed to reduce the number of continuation parameters by introducing an extra Poincar\'e section. This gives us an applicable inspiration to deal with the computation of the global invariant manifold in the three degree of freedom system as well. The current study on the linearized dynamics also presents a possible way, that is, to first find the boundary of the topological tube and ellipsoid in the position space and then determine the cone of velocity at each admissible point. In  summary,  future work can  reveal the geometry of escape dynamics in higher degree of freedom systems or develop corresponding computational algorithms for the global invariant manifolds, extending the local picture given here.

\section{Conclusions}
\label{conclusions}
In this paper, we study the escape dynamics in a three degree of freedom spring-mass system in the presence of energy dissipation, presenting both dynamical structures and transition criteria. This study only focuses on the local dynamics around the neck region of the index-1 saddle. From the analytical derivation, we found that the type of the equilibrium point changes from a
saddle $\times$ center $\times$ center in the conservative system to saddle $\times$ focus $\times$ focus in the dissipative system. 
Furthermore, we demonstrated that the transition boundaries for the conservative and dissipative systems of a given energy $h$ are the stable invariant manifolds of the corresponding center manifold at the energy  $h$ and the equilibrium point, respectively, which appear as a 4-dimensional hyper-cylinder and a 4-dimensional hyper-ellipsoid, respectively. The orbits with initial conditions inside of the transition boundary are transit orbits, while those with initial conditions outside of the transition boundary are non-transit orbits. In the position space, the initial conditions should satisfy two criteria: (1) the initial position should be inside of the tube of transition and the ellipsoid of transition for the conservative and dissipative systems, respectively, and (2) the velocity should be interior to a cone of velocity at that position.

Although this paper investigated the escape dynamics in a three degree of freedom system taking into account the dissipative forces, only linear behaviors around the equilibrium region were presented. When a larger energy is input to the system, the nonlinear behaviors become prominent. 
The discussion of the full nonlinear system is left as future work to reveal the topological phase space structures that govern the transition in three and higher degrees of freedom system from a global perspective. 
Based on the conclusions in the current study, one possible avenue for computing the global transition boundary is to extend the current local invariant manifold method to the corresponding global invariant manifold \cite{zhong2021global}.

\section*{Acknowledgements}
This work was partially supported in part by the National Science Foundation under awards 1922516 and 2027523 to SDR.


\section*{References}


\renewcommand\refname{ References} 
\bibliographystyle{shane-unsrt} 

\bibliography{Jun}
\end{document}